\documentclass[a4paper,12pt]{article}
\usepackage[left=25mm, right=25mm, bottom=35mm, top=35mm]{geometry}

\usepackage[T2A]{fontenc}
\usepackage[utf8]{inputenc}
\usepackage[english]{babel}
\usepackage[pdftex]{graphicx}

\usepackage{amsmath,amssymb,ntheorem}
\usepackage{multicol}
\usepackage{colortbl} %%-added for color tables

\usepackage{geometry}
\usepackage{epsf}
\usepackage{amsfonts}
\usepackage{amsmath}
\usepackage{amssymb}
\usepackage{latexsym}
\usepackage{units}
\usepackage{textcomp}
\usepackage{wrapfig}
\usepackage{booktabs}
\usepackage{hyperref}
\usepackage{xcolor}
\hypersetup{
    colorlinks,
    linkcolor={red!50!black},
    citecolor={blue!50!black},
    urlcolor={blue!80!black}
}

\everymath{\displaystyle}
\usepackage{color}
\everymath{\displaystyle}

\renewcommand{\phi}{\varphi}

\usepackage{url}

\begin{document}

\begin{center}
\title{\boldmath Measurement of the $K^+\rightarrow{\mu^+}{\nu_{\mu}}{\gamma}$ 
decay form factors in the OKA experiment}
\section*{ \boldmath  Measurement of the $K^+\rightarrow{\mu^+}{\nu_{\mu}}{\gamma}$ decay form factors in the OKA experiment}
\end{center}

\begin{center}
{\large 
\textsc{The OKA collaboration}\\
}
\vspace{3mm}

\begin{minipage}{0.9\linewidth}
 \center{
  %\small
  \rmfamily
  V.I.~Kravtsov, V.A.~Duk\footnote{\scriptsize {\it Present Address}:~University~of~Birmingham,~Birmingham,~United~Kingdom}, 
%  V.I.~Kravtsov, V.A.~Duk\footnote{\scriptsize Also~at~University~of~Birmingham,~Birmingham,~United~Kingdom}, 
  S.N.~Filippov, E.N.~Gushchin, A.A.~Khudyakov,  
  Yu.G.~Kudenko\footnote{\scriptsize Also at Moscow Institute of Physics and Technology, Moscow Region, Russia}\footnote{\scriptsize Also at NRNU Moscow Engineering Physics Institute (MEPhI), Moscow, Russia}, 
  A.Yu.~Polyarush
 }\vspace{-4mm}
 \center{\small 
   %\itshape 
    \textsc{(INR RAS, Moscow, Russia),}
 }\\
 \vspace{-3mm}
\center{
  %\small 
  %\ttfamily
  %\rmfamily
  %\textsf
  \textsc
  S.A.~Akimenko, A.V.~Artamonov, A.M.~Blik, V.S.~Burtovoy, 
  S.V.~Donskov, A.P.~Filin, A.M.~Gorin, A.V.~Inyakin, 
  G.V.~Khaustov, S.A.~Kholodenko, V.N.~Kolosov, V.F.~Kurshetsov, 
  V.A.~Lishin,  M.V.~Medynsky, Yu.V.~Mikhailov, V.F.~Obraztsov, 
  V.A.~Polyakov, A.V.~Popov, V.I.~Romanovsky, V.I.~Rykalin, A.S.~Sadovsky, V.D.~Samoilenko, 
  V.K.~Semenov\footnote{Deceased}, M.M.~Shapkin, O.V.~Stenyakin,  O.G.~Tchikilev, V.A.~Uvarov, O.P.~Yushchenko
 }
 \vspace{-4mm}
 \center{\small 
   %\itshape
   \textsc{(NRC "Kurchatov Institute"${}^{}_{}{}^{}$-${}^{}_{}{}^{}$IHEP, Protvino, Russia),} 
 }\\
 \vspace{-3mm}
 \center{
  %\small
  \rmfamily
  V.N.~Bychkov, G.D.~Kekelidze, V.M.~Lysan, B.Zh.~Zalikhanov
 }\vspace{-4mm}
 \center{\small 
   \itshape 
   \textsc{(JINR, Dubna, Russia)}\\
 }
\end{minipage}
\end{center}

\vspace{0mm}
\begin{center}
\begin{minipage}{0.09\linewidth}
~
\end{minipage} 

\begin{minipage}{0.80\linewidth}
{ \rmfamily

{\bf Abstract.}
A precise measurement of the vector and axial-vector form factors difference 
$F_V-F_A$ in the $K^+\rightarrow{\mu^+}{\nu_{\mu}}{\gamma}$ decay is presented. 
About 95K events of $K^+\rightarrow{\mu^+}{\nu_{\mu}}{\gamma}$ are selected in the 
OKA experiment. The result is $F_V-F_A=0.134\pm0.021(stat)\pm0.027(syst)$. 
Both errors are smaller than in the previous $F_V-F_A$ measurements. 
}
\end{minipage}

\begin{minipage}{0.09\linewidth}
~
\end{minipage}
\end{center}
\vspace{0mm}

\selectlanguage{english}

\section{Introduction}
Radiative kaon decays are sensitive to hadronic weak currents in low-energy 
region and provide a good testing for the chiral perturbation theory 
(${\chi}PT$). The amplitude of the $K^+\rightarrow{\mu^+}{\nu_{\mu}}{\gamma}$ 
decay includes two terms: internal bremsstrahlung (IB) and structure dependent 
term (SD) \cite{neville}. IB contains radiative corrections for 
$K^+\rightarrow{\mu^+}{\nu_{\mu}}$ decay. SD is sensitive to the electroweak 
structure of the kaon. 

The differential decay rate can be written in terms of standard kinematic 
variables \\
$x=2E^{\ast}_{\gamma}/M_K$ and $y=2E^{\ast}_{\mu}/M_K$ \cite{bijnens}, which 
are proportional to the photon $E^{\ast}_{\gamma}$ and muon $E^{\ast}_{\mu}$ 
energy in the kaon rest frame ($M_K$ is the kaon mass). It includes IB, 
SD$^\pm$ terms and their interference INT$^\pm$. The SD$^\pm$ and INT$^\pm$ 
contributions are determined by two form factors $F_V$ and $F_A$.

The general formula for the decay rate is as follows: \\
$\frac{d\Gamma}{dxdy}=A_{IB}f_{IB}(x,y)+A_{SD}[(F_V+F_A )^2 
f_{SD^+}(x,y)+(F_V-F_A )^2 f_{SD^-}(x,y)]$ \\
\hspace*{11 mm}$-A_{INT}[(F_V+F_A )f_{INT^+}(x,y)+(F_V-F_A )f_{INT^-}(x,y)]$, \\
where \\
$f_{IB}(x,y)=[\frac{1-y+r}{x^2(x+y-1-r)}][x^2+2(1-x)(1-r)
-\frac{2xr(1-r)}{x+y-1-r}]$, \\
$f_{SD^+}(x,y)=[x+y-1-r][(1-x)(1-y)+r]$, \\
$f_{SD^-}(x,y)=[1-y+r][(x+y-1)(1-x)-r]$, \\
$f_{INT^+}(x,y)=[\frac{1-y+r}{x(x+y-1-r)}][(1-x)(1-x-y)+r]$, \\
$f_{INT^-}(x,y)=[\frac{1-y+r}{x(x+y-1-r)}][x^2-(1-x)(1-x-y)-r]$, \\
and $r=[\frac{M_{\mu}}{M_K}]^2$, 
$A_{IB}=\Gamma_{K_{\mu2}}\frac{\alpha}{2\pi}\frac{1}{(1-r)^2}$, 
$A_{SD}=\Gamma_{K_{\mu2}}\frac{\alpha}{8\pi}\frac{1}{r(1-r)^2}[\frac{M_K}{F_K}]^2$, \\
$A_{INT}=\Gamma_{K_{\mu2}}\frac{\alpha}{2\pi}\frac{1}{(1-r)^2}[\frac{M_K}{F_K}]$. 
Here $\alpha$ is the fine structure constant, $F_K$ is $K^+$ decay constant 
($F_K = 155.6\pm0.4\:MeV$ \cite{nakamura}) and $\Gamma_{K_{\mu2}}$ 
is the $K_{\mu2}$ decay width.

\begin{figure}[h]
\centering
\includegraphics[width=0.95\textwidth]{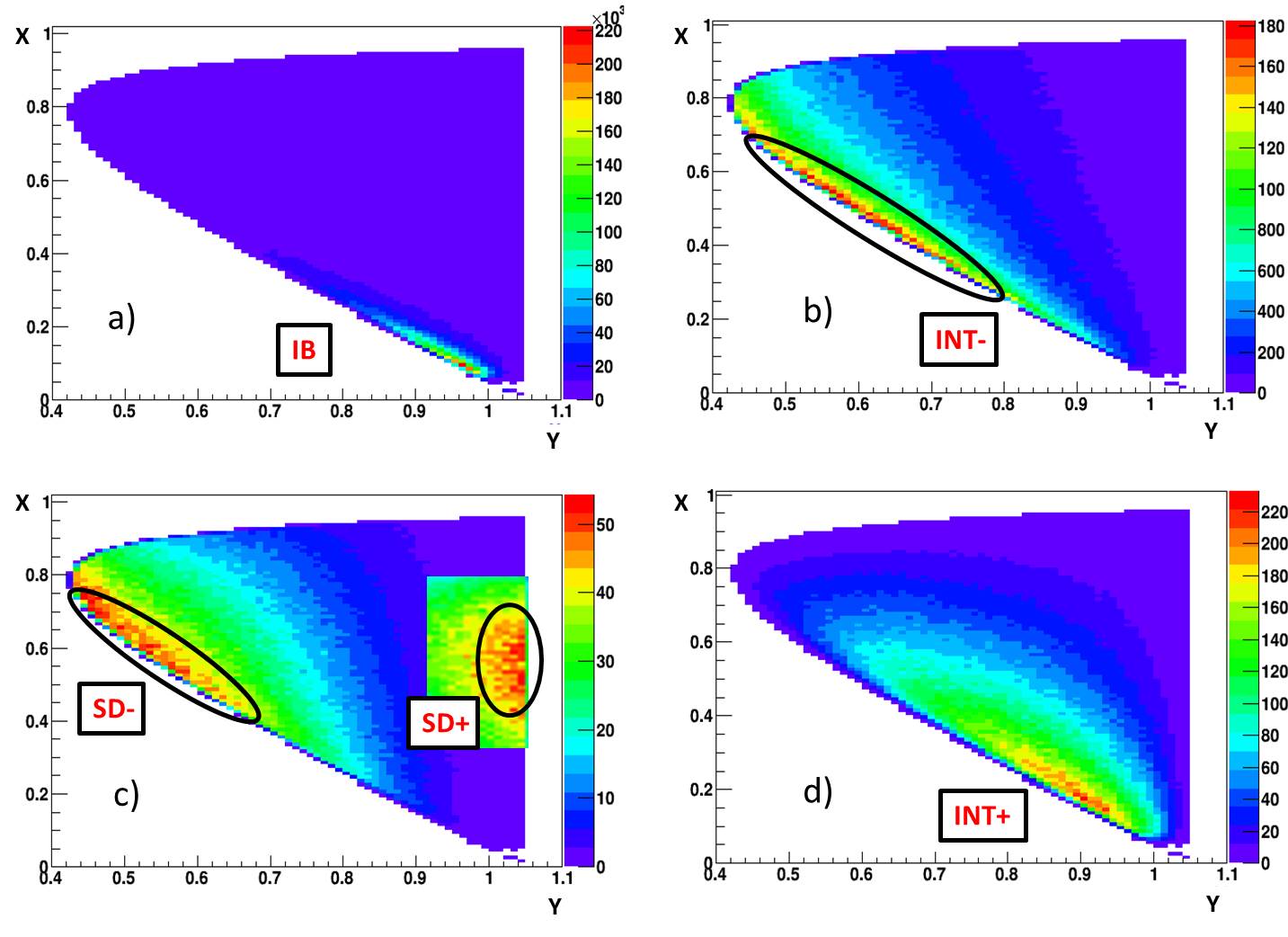}
\caption{\label{fig:kmng5}Kinematic distribution of different terms: (a) IB, 
(b) INT$^-$, (c) SD$^-$ and SD$^+$, (d) INT$^+$.}
\end{figure}

Fig.~\ref{fig:kmng5} shows the kinematic distribution for 
IB, INT$^-$, INT$^+$, SD$^-$ and SD$^+$. 
The main goal of the analysis is to measure ${F_V}-{F_A}$ by extracting the 
INT$^-$ term. Other terms are either suppressed by backgrounds or give negligible 
contribution to the total decay rate with respect to IB. In the lowest order of 
${\chi}PT\:O(p^4)$ $F_V$ and $F_A$ are constant and ${F_V}-{F_A}=0.052$ 
\cite{bijnens}. The first measurement of ${F_V}-{F_A}$ was made by the ISTRA+ 
experiment: $F_V-F_A=0.21{\pm}0.04(stat){\pm}0.04(syst)$ \cite{istra}. 

\section{OKA detector and separated kaon beam}
\begin{figure}[h]
\centering
\includegraphics[width=1.0\textwidth]{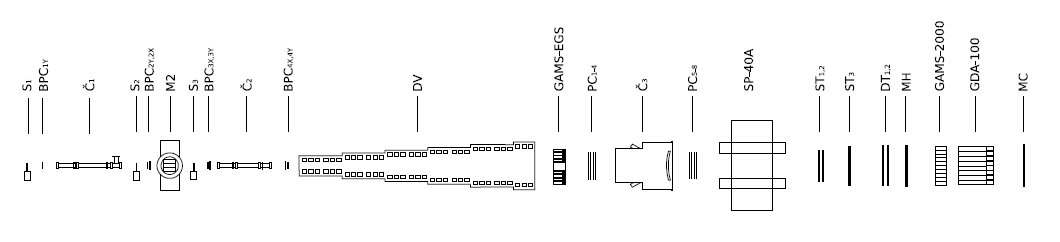}
\caption{OKA setup. The particle beam goes from left to right.}
\label{fig:OKA_detector_old}
\end{figure}

The OKA setup, Fig.~\ref{fig:OKA_detector_old}, is a double magnetic spectrometer. 

The OKA detector includes: 
\begin{itemize}
\item Beam spectrometer consisting of the magnet M2, 7 beam proportional chambers 
BPC, 4 beam scintillation counters S and 2 threshold Cherenkov counters 
$\check{C}_{1,2}$ for the kaon identification; 
\item 11 m long He filled decay volume DV with the guard system (GS) 
containing 670 Lead-Scintillator calorimetric modules 
20$\times$(5 mm Sc + 1.5 mm Pb) with WLS readout; 
\item Main magnetic spectrometer on the basis of $200{\times}140$ cm$^2$ wide 
aperture magnet SP-40A with a field integral 1 Tm, complemented by 13 planes of 
proportional chambers (PC), straw (ST) and drift tubes (DT); 
\item 2 gamma detectors: electromagnetic calorimeter GAMS-2000 and large angle 
detector EGS (EGS is used to supplement GS as a gamma veto at large angles); 
\item Hadron calorimeter GDA-100 and 4 muon scintillation counters $\mu$C 
(marked as MC in Fig.~\ref{fig:OKA_detector_old}) used for muon identification; 
\item Pad (Matrix) Hodoscope MH for the trigger and track reconstruction. 
\end{itemize}
More details can be found in \cite{sadovsky}. 

The data acquisition system of the OKA setup \cite{daq} operates
at $\sim{25}$ kHz event rate with the mean event size of $\sim{4}$ kByte. 

The OKA beam is a separated secondary beam of the U-70 Proton Synchrotron 
of NRC "Kurchatov Institute"-IHEP, Protvino \cite{beam}. RF-separation with the Panofsky scheme 
\cite{panofsky} is implemented. The beam contains up to $12.5\%$ of kaons with an 
intensity of about $5\times{10^5}$ kaons per 3 sec U-70 spill. The beam momentum was 
17.7 GeV/c during the data taking period used for the analysis (November 2012). 
The present study uses about 1/2 of the statistics collected in 2012, where 
504M events were stored on tape. 

\section{Trigger streams and primary selection}
The following trigger was used for the analysis: 
$T_{GAMS}=beam\cdotp\overline{\check{C_1}}\cdotp\check{C_2}\cdotp
\overline{S_{bk}}\cdotp{E_{GAMS}}$, where 
$beam={S_1}\cdotp{S_2}\cdotp{S_3}\cdotp{S_4}$ is a coincidence of four beam 
scintillation counters, $\check{C}_{1,2}$ - threshold Cherenkov counters 
($\check{C_1}$ selects pions, $\check{C_2}$ - pions and kaons), 
$S_{bk}$ (''beam killer'') - two scintillation counters on the beam axis after the 
magnet aimed to suppress undecayed beam particles. The analog amplitude sum in 
the GAMS-2000 is required be higher than $E_{GAMS}$ ($E_{GAMS}$ is chosen to be 
above the average MIP energy deposit). The 10 times prescaled minimum bias trigger 
$T_{kaon}=beam\cdotp\overline{\check{C_1}}\cdotp\check{C_2}\cdotp\overline{S_{bk}}$ 
was used for the trigger efficiency measurement 
$\epsilon_{trig}={(T_{GAMS}\cap{T_{kaon}})}/{T_{kaon}}$ 
(Fig.~\ref{fig:trig_efficiency}). 
This trigger efficiency was applied during the Monte Carlo (MC) simulation.  

To select the decay channel the following requirements are applied: 
\begin{itemize}
\item 1 primary track; 
\item 1 secondary track identified as muon in GAMS-2000, GDA-100 and $\mu$C; 
\item 1 electromagnetic shower in GAMS-2000 with energy $E_{tot}>1\:GeV$ not 
associated with charged track; 
\item GS energy deposition $E_{GS}<10\:MeV$; 
\item EGS energy deposition $E_{EGS}< 100\:MeV$; 
\item Decay vertex inside the decay volume DV. 
\end{itemize}

\section{Event selection}
The main background to the $K^+\rightarrow{\mu^+}{\nu_{\mu}}{\gamma}$ decay 
comes from 2 decay modes: $K^+\rightarrow{\mu^+}{\nu_{\mu}}{\pi^0}$ ($K\mu3$) 
and $K^+\rightarrow{\pi^+}{\pi^0}$ ($K2\pi$) with one $\gamma$ lost from 
$\pi^0\rightarrow{\gamma\gamma}$ decay and $\pi$ misidentified as $\mu$. 
Additional contribution at $y>1$ is given by the decay mode 
$K^+\rightarrow{\mu^+}{\nu_{\mu}}$ with an accidental $\gamma$. 
At low y values there is a small contribution from the 
$K^+\rightarrow{\pi^+}{\pi^-}{\pi^+}$ ($K3\pi$) decay. 

The MC simulation of the OKA setup is done within the GEANT3 framework 
\cite{geant3}. Signal and background events are weighted according to 
corresponding matrix elements. 

The $K^+\rightarrow{\mu^+}{\nu_{\mu}}{\gamma}$ event selection strategy is based 
on the ISTRA+ approach \cite{istra}. Signal extraction procedure starts with 
dividing all kinematic ($x,\:y$) region into strips in $x$ with $\Delta{x}=0.05$ 
width. The following steps are implemented for each $x$-strip: 
\begin{itemize}
\item Fill the $y$ plot.
\item Select the signal region by a cut $y_{min}<y<y_{max}$ and fill 
$\cos{\theta^{\ast}_{\mu\gamma}}$ plot, where $\theta^{\ast}_{\mu\gamma}$ is an 
angle between $\mu$ and $\gamma$ in the kaon rest frame. $y_{min}$ and $y_{max}$ 
are selected from the maximization of signal significance defined as 
$S/\sqrt{S+B}$ where $S$ is the signal and $B$ is the background. 
\item Put a cut on $\cos{\theta^{\ast}_{\mu\gamma}}$ to reject background and fill 
$m_k$ plot. $m^{2}_{k}=(P_{\mu}+P_{\nu}+P_{\gamma})^2$, where 
$P_{\mu}$,$\:P_{\nu}$,$\:P_{\gamma}$ are 4-momenta of decay particles in the laboratory 
frame, \\
$\vec{p}_{\nu}=\vec{p}_{K}-\vec{p}_{\mu}-\vec{p}_{\gamma}$, 
$E_{\nu}=|\vec{p}_{\nu}|$. $m_k$ peaks at the kaon mass for the signal. 
\item The last step is a simultaneous fit of all 3 histograms 
($y, \cos{\theta^{\ast}_{\mu\gamma}}, m_k$) with the \\
MINUIT tool \cite{minuit} 
where the signal and backgrounds normalization factors are the fit parameters. 
\end{itemize}

For the correct estimation of the statistical error $\sigma_{exp}$, only 
the $m_k$ histogram is used. The MINOS program \cite{minuit} is run once with 
the initial parameter values equal to those obtained in the simultaneous fit. 
Statistical errors were extracted from the MINOS output. 

\begin{figure}[h]
\begin{minipage}[t]{0.47\linewidth}
\includegraphics[width=0.97\textwidth]{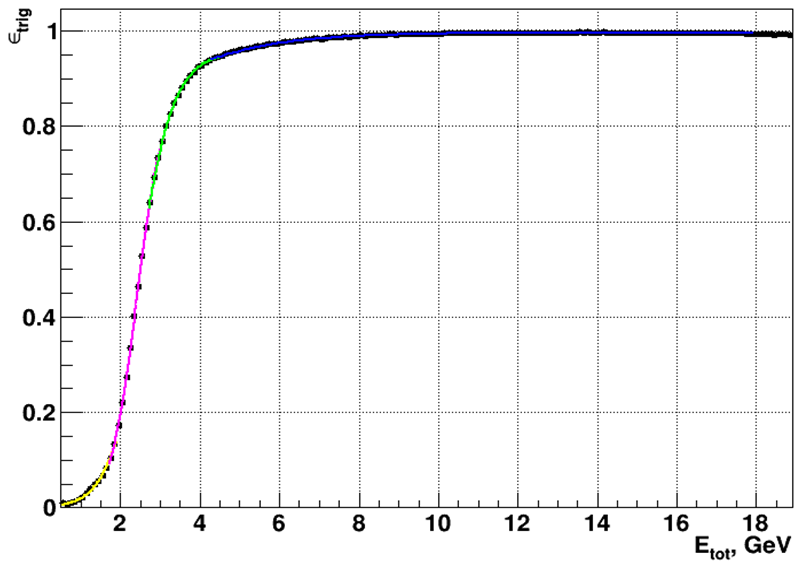}
\caption{Trigger efficiency 
$\epsilon_{trig}$ as the function of the GAMS total energy deposition. 
Black points - data, colored curves - fit by the third degree polynomial 
in four intervals.}
\label{fig:trig_efficiency}
\end{minipage}\hspace{2pc}%
\begin{minipage}[t]{0.47\linewidth}
\includegraphics[width=1.0\textwidth]{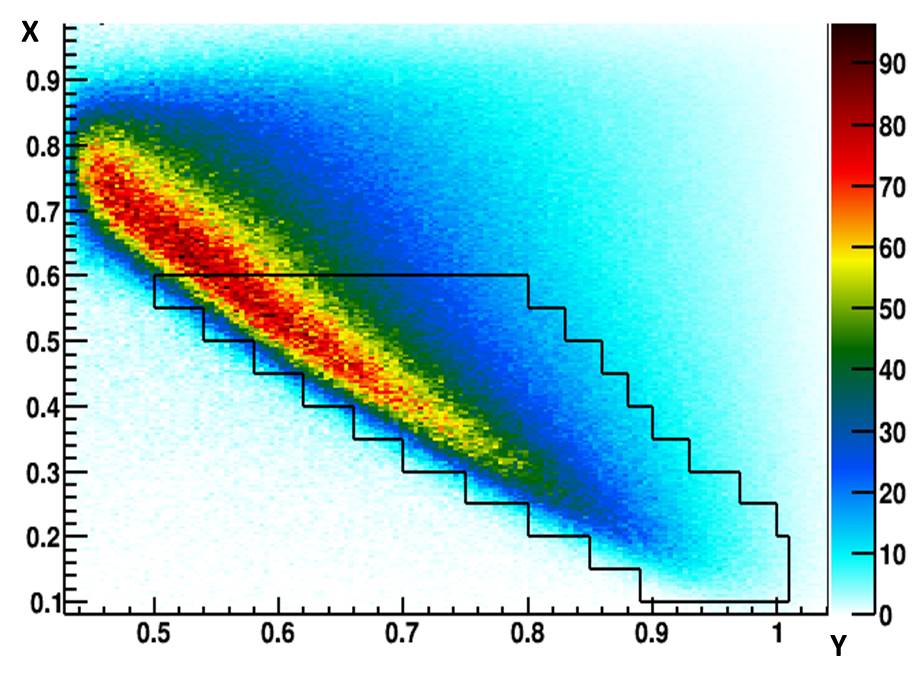}
\caption{INT$^-$ Dalitz-plot density and selected kinematic region 
(area contoured by the black line).}
\label{fig:sel_region}
\end{minipage}
\end{figure}

Fig.~\ref{fig:sel_region} shows the selected kinematic region for the extraction 
of the INT$^-$ term. For the further analysis 10 $x$-strips were selected in 
the $0.1<x<0.6$ region. 
The $y$-width varies from 0.12 to 0.30 inside $x$-strips. 

\begin{figure}[h]
\centering
\includegraphics[width=0.95\textwidth]{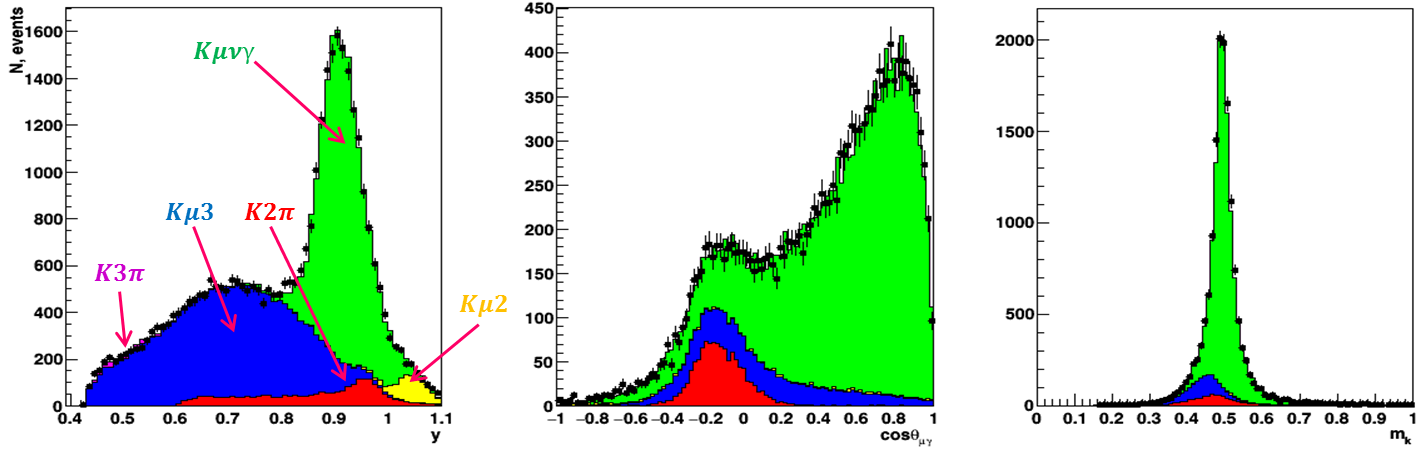}
\caption{Simultaneous fit in the strip 2 (0.15 $<$ x $<$ 0.20): 
y, $\cos{\theta^{\ast}_{\mu\gamma}}$, $m_k$. Black points with errors - data, 
blue - $K\mu3$, red - $K2\pi$, yellow - $K\mu2$, violet - $K3\pi$, 
green - signal. ${\chi^2}/NDF=1.7$.}
\label{fig:strips_2}
\end{figure}

The result of the simultaneous fit for the strip 2 ($0.15<x<0.2$) is shown in 
Fig.~\ref{fig:strips_2}. Both signal and background shapes are taken from 
the MC simulation. 
The total normalization of the MC to data is made to the $K\mu3$ decay at 
$y<0.6$, where the contribution of other backgrounds is very small. The relative 
normalization of other backgrounds is done according to their 
branching ratios.
For the $K^+\rightarrow{\mu^+}{\nu_{\mu}}{\gamma}$ decay, only IB term is 
included in the simultaneous fit. The simultaneous fit gives a reasonable 
agreement between data and MC with $\chi^2$ from 1.3 to 1.7 for different 
$x$-strips. 

\section{$F_V-F_A$ calculation}
\begin{figure}[h]
\centering
\includegraphics[width=0.6\textwidth]{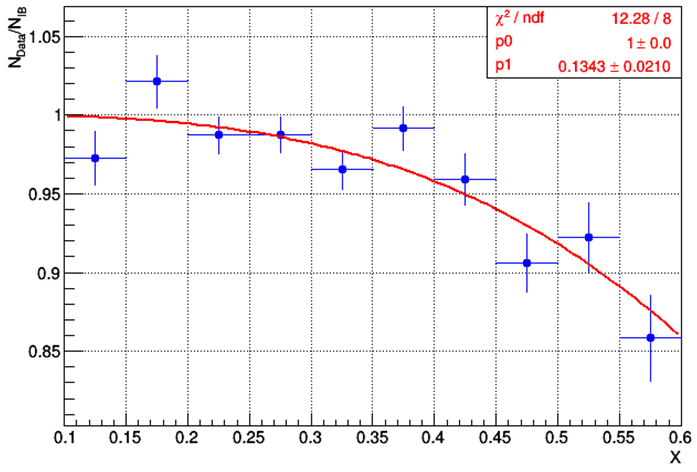}
\caption{$N_{Data}/N_{IB}$ ratio as a function of $x$ (blue points with 
errors) and result of the fit with $p_{signal}(x)$ (red line). For the definition 
of $p_{signal}(x)$, see text.}
\label{fig:result_fit}
\end{figure}

For each $x$-strip the number of signal events $N_{Data}$ is extracted from 
the simultaneous fit and the IB event number $N_{IB}$ is obtained from MC. 
Their ratio is plotted as a function of $x$ (Fig.~\ref{fig:result_fit}). 
For the signal containing IB only this ratio would be equal to 1. It is the case 
for small $x$, when the IB is dominating and INT$^-$ is negligible. For large $x$ the 
INT$^-$ term gives significant negative contribution resulting in smaller values of 
$N_{Data}/N_{IB}$. \\
The $N_{Data}/N_{IB}$ distribution is fitted with a function 
$p_{signal}(x)=p_{0}(1+p_{1}({{\phi}_{INT^-}}(x)/{{\phi}_{IB}}(x)))$, 
where $p_0$ is normalization factor, 
$p_1=F_V-F_A$ is the difference of vector and axial-vector form factors, 
${{\phi}_{INT^-}}(x)$ is the $x$-distribution for the reconstructed MC-signal events 
taken with the weights $({M_K}/{F_K}){f_{INT^-}}(x_{true},y_{true})$, 
${{\phi}_{IB}}(x)$ is a similar distribution for the same MC sample, but with the 
weights ${f_{IB}}(x_{true},y_{true})$. 
Here $x_{true}$, $y_{true}$ are "true" MC values of $x$ and $y$. 

The result of the fit is $F_V-F_A=0.134{\pm}0.021$. The normalization factor is 
$p_{0}=1.000{\pm}0.007$. The total number of selected 
$K^+\rightarrow{\mu^+}{\nu_{\mu}}{\gamma}$ decay events is $95428\pm309$. 

\begin{figure}[h]
%\begin{minipage}{14pc}
\begin{minipage}[t]{0.47\linewidth}
\includegraphics[width=1.0\linewidth]{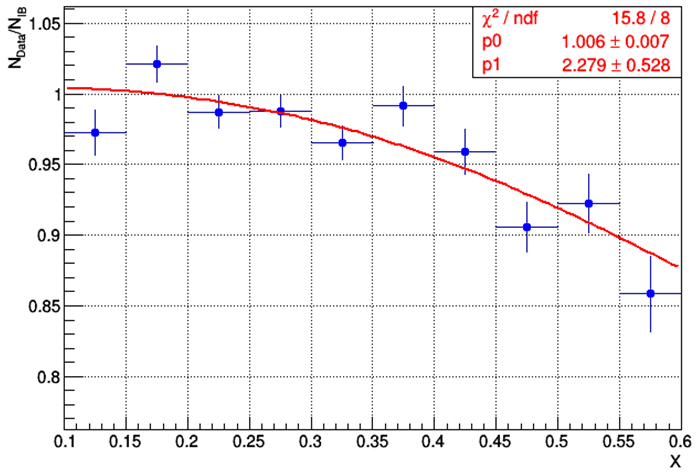}
\caption{\label{label}${\chi}PT\:O(p^6)$ fit, $F_V(0)$ and $F_A$ are taken 
from theory. The fit gives $\lambda=2.279\pm0.528$.}
\label{fig:op6_lambda}
\end{minipage}\hspace{2pc}
\begin{minipage}[t]{0.47\linewidth}
\includegraphics[width=1.0\linewidth]{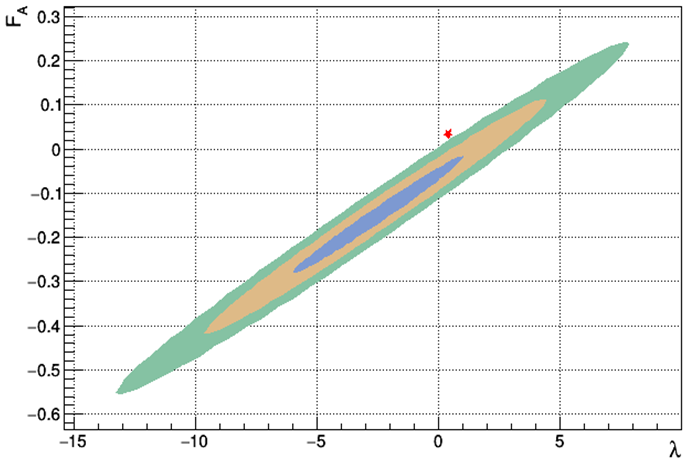}
\caption{\label{label}($F_V,\:\lambda$) correlation plot. $F_V(0)$ is taken 
from theory. Red star is theory prediction.}
\label{fig:op6_correlation}
\end{minipage} 
\end{figure}

In the next order ${\chi}PT\:O(p^6)$ $F_V$ linearly depends on the momentum 
transfer $q^2$ \cite{geng} with the following parametrization 
\cite{ambrosino}: ${F_V}={F_V(0)}(1+{\lambda}(1-x))$, ${F_A}=const$. 
The theoretical prediction is tested in three ways: 
\begin{itemize}
\item The final fit is performed with $F_V$ and $F_A$ fixed from 
${\chi}PT\:O(p^6)$ prediction:  $F_V(0)=0.082$, $F_A=0.034$, ${\lambda}=0.4$. 
This fit has bad compliance with ${\chi^2}/NDF=28.0/9$. 
\item $F_V(0)$ and $F_A=0.034$ are taken from ${\chi}PT\:O(p^6)$, $\lambda$ 
is a fit parameter. It gives ${\lambda}=2.28\pm0.53$ with ${\chi^2}/NDF=15.8/8$ 
(Fig.~\ref{fig:op6_lambda}).
\item $F_V(0)$ is fixed from ${\chi}PT\:O(p^6)$. $F_A$ and $\lambda$ are used 
as fit parameters. ($F_V,\:\lambda$) correlation is shown in 
Fig.~\ref{fig:op6_correlation}. The theoretical prediction (red star) is slightly 
out of $3\sigma$-ellipse. 
\end{itemize}

\section{Systematic errors}
The obtained value of $F_V-F_A$ depends on the width of $x$-strips, $y$ and 
$\theta^{\ast}_{\mu\gamma}$ cuts and the fit procedure. 
The following sources of the systematic errors were investigated: 
\begin{itemize}
\item Non ideal description of signal and background by the MC. \\
For the estimation of this systematics, the statistical error in each bin of 
Fig.~\ref{fig:result_fit} 
was scaled by the factor $\sqrt{{\chi^2}/NDF}$, where ${\chi^2}/NDF$ is obtained 
from the simultaneous fit in each $x$-strip. 
A new fit of $N_{Data}/N_{IB}$ with the same function $p_{signal}(x)$ gives 
the best description with ${\chi^2}/NDF=7.8/8$ compared to ${\chi^2}/NDF=12.3/8$ 
of the main fit. The new value of $F_V-F_A=0.138$ is consistent with the main one 
but the fit error $\sigma_{fit}=0.026$ is larger. Assuming 
${\sigma_{fit}}^2={\sigma_{shape}}^2+{\sigma_{stat}}^2$ the systematic error is 
$\sigma_{shape}=0.015$. 
\item The fit range in $x$ (number of $x$-strips in the fit). \\
The ratio $N_{Data}/N_{IB}$ was refitted by removing one or two bins on the 
left (right) edge. For the estimate of systematics the average difference between 
the new $F_V-F_A$ values and the nominal one is taken. 
The error is negligible: $\sigma_{x}<0.006$. 
\item Width of $x$-strips. \\
The $F_V-F_A$ calculation is repeated for 2 different values of $x$-binning: 
$\Delta{x}=0.035$, $\Delta{x}=0.07$. 
The deviation of the new $F_V-F_A$ value with respect to the main one gives 
$\sigma_{\Delta{x}}=0.011$. 
\item $y$ limits in $x$-strips. \\
The events inside FWHM of the $y$-distribution for the signal MC are selected.  
Such limits are tighter than those used in the main analysis. 
The difference between the new value and main one gives systematic 
error $\sigma_{y}=0.008$. 
\item Possible contribution of INT$^+$. \\
The INT$^+$ term is added to the final fit (see Section 5). The value 
$|F_V+F_A|=0.165{\pm}0.013$ measured by the E787 experiment is used \cite{adler}. 
Two fits were repeated for the minimal (-0.178) and maximal (+0.178) possible 
values of $F_V+F_A$. 
The fitting function was: \\
$p_{signal}(x)=p_{0}(1+({F_V+F_A})({{\phi}_{INT^+}}(x)/{{\phi}_{IB}}(x))+
({F_V-F_A})({{\phi}_{INT^-}}(x)/{{\phi}_{IB}}(x)))$, \\
where ${{\phi}_{INT^+}}(x)$ is the $x$-distribution similar one as 
${{\phi}_{INT^-}}(x)$. 
The maximal difference between obtained values of $F_V-F_A$ and the main one 
measured in Section 5 is $\sigma_{INT^{+}}=0.018$.
\end{itemize}
Summing up quadratically all the systematic errors the total error is found 
to be 0.027. 

\section{Conclusion}
The largest statistics of about 95K events of 
$K^+\rightarrow{\mu^+}{\nu_{\mu}}{\gamma}$ is collected by the OKA experiment. 
The INT$^-$ term is observed and $F_V-F_A$ is measured: \\
$F_V-F_A=0.134{\pm}0.021(stat){\pm}0.027(syst)$. \\
The result is $2.4\sigma$ above ${\chi}PT\:O(p^4)$ prediction. 

A recent calculation in the framework of the gauged nonlocal effective chiral 
action ($E{\chi}A$) gives $F_V-F_A=0.081$ \cite{shim}. 
The OKA result is $1.6\sigma$ above the $E{\chi}A$ prediction. 

The obtained value of $F_V-F_A$ is in a reasonable agreement with a similar analysis 
of the ISTRA+ experiment: 
$F_V-F_A=0.21{\pm}0.04(stat){\pm}0.04(syst)$ \cite{istra}. 
%and the alternative ISTRA+ analysis \\
%($F_V-F_A=0.126{\pm}0.027(stat){\pm}0.043(syst)$) \cite{chikilev} 
%with reduced errors. 

\section{Acknowledgements}
The authors express their gratitude to IHEP colleagues from the accelerator department 
for the good performance of the U-70 during data taking; from the beam department 
for the stable operation of the 21K beam line, including RF-deflectors, and from 
the engineering physics department for the operation of the cryogenic system 
of the RF-deflectors. 

The work is partially supported  by the Russian Fund for Basic Research, \\
grant N18-02-00179A.

\end{document}